\def\glt{\mbox{\hspace{0.4em}} {}^{>} \mbox{\hspace{-0.6em}}
  {}_{<} \mbox{\hspace{0.4em}}}

\documentstyle[a4,12pt,epsf,subfigure,wrapfloat]{article}

\begin{document}

\vspace*{-3em}
\begin{flushright}
  DPNU-96-06 \\ (Revised) \\ January 1996
\end{flushright}

\begin{center}
  \setlength{\baselineskip}{20pt}
  {\LARGE Slow Relaxation at Critical Point of Second Order Phase
    Transition in a Highly Chaotic Hamiltonian System}\\
  \ \\
  {\large Yoshiyuki Y. YAMAGUCHI }
  \footnote{e-mail: yamaguchi@allegro.phys.nagoya-u.ac.jp}\\
  {\large Department of Physics, School of Science,\\
    Nagoya University, Nagoya, 464-01, Japan}
\end{center}

\setlength{\baselineskip}{15pt}

\begin{abstract}
Temporal evolutions toward thermal equilibria are numerically
investigated in a Hamiltonian system with many degrees of freedom
which has second order phase transition. Relaxation processes are
studied through local order parameter, and slow relaxations of power
type are observed at the critical energy of phase transition for some
initial conditions. Numerical results are compared with results of a
phenomenological theory of statistical mechanics. At the critical
energy, the maximum Lyapunov exponent takes the largest value.
Temporal evolutions and probability distributions of local Lyapunov
exponents show that the system is highly chaotic rather than weakly
chaotic at the critical energy. Consequently theories for perturbed
systems may not be applied to the system at the critical energy in
order to explain the slow relaxation of power type.
\end{abstract}

\section{Introduction}
\label{sec:introduction}

In recent years some researchers have investigated Hamiltonian
systems with many degrees of freedom \cite{flach-94} (and references
therein). They are interested in statistical mechanics since
Hamiltonian dynamics is the basis of it. They mainly have considered
relaxation processes to equilibrium states because we can observe
temporal evolutions of systems directly from Hamiltonian dynamics.

In both the one-dimensional classical $\Phi^4$ lattice model and FPU
$\beta$ model, they observed slow and fast relaxation processes in low
and high energy regions, respectively
\cite{pettini-90}\cite{flach-93}. The slow and fast relaxation 
processes occur in weakly and highly chaotic systems, respectively 
\cite{mutschke-93}. Here the word of weakly chaotic system means that
theories for perturbed systems may be applied for the system. For
instance, if we add small non-integrable perturbation to a completely
integrable system, whose phase space consist of tori, almost all tori
survive under the small perturbation. That is a result of
Kolmogorov-Arnold-Moser (KAM) theorem \cite{kam}. Intermittency
between chaotic and regular-like motions is also a characteristic
phenomenon which occurs in perturbed systems. Arnold diffusion
,\cite{chirikov-79} Nekhoroshev time\cite{nekhoroshev-77} and other
theories exist for perturbed systems. 
Using theories and properties of weakly chaotic systems,
above-mentioned slow relaxation, which occurs in weakly chaotic
systems, was discussed\cite{pettini-90}.

Properties of weakly chaotic systems have been well investigated both
theoretically and numerically. Contrary to weakly chaotic systems, few
studies exist for highly chaotic systems, and to study properties of
highly chaotic systems whose dynamics can not reduced to Markovian
process is a left important problem of classical Hamiltonian systems.

We expect it is useful to consider relaxation processes in a
Hamiltonian system having second order phase transition in order to
study properties of highly chaotic systems. The reason is as follows.
The system seems highly chaotic near the critical point because
fluctuation anomalously increases. Nevertheless, according to a
phenomenological theory of statistical mechanics, van Hove theory,
slow relaxation of power type of order parameter appears at the
critical point. That means a certain structure exists in phase space.

In this way we understand that critical phenomenon is interesting as a
problem in dynamics, as well as in statistical mechanics.

In this paper, we investigate the following things. First, as a 
problem of statistical mechanics, we verify appearance of the slow
relaxation of order parameter at the critical point even when we use a
dynamical method, and we compare results of dynamics with ones of
phenomenological theory. Second, as a problem of dynamics, we confirm
that the system is highly chaotic at the critical point, and we
suggest that critical phenomenon is a useful phenomenon to study
properties of highly chaotic systems. 

We introduce a Hamiltonian system in which second order phase
transition occurs, and numerically integrate the equations of motion.
The model Hamiltonian is as follows \cite{ruffo-94}
\begin{equation}
  H(q,p)=\sum^N_{i=1}\frac{1}{2}p_i^2 + U(q),
  \label{Hamiltonian}
\end{equation}
\begin{equation}
  \label{potential}
  U(q) = \frac{1}{2N}\sum^N_{i,j=1} (1-\cos(q_i-q_j)).
\end{equation}
Equations of motion are derived from the Hamiltonian as follows
\begin{eqnarray}
  \frac{dq_i}{dt} &=& p_i \\
  \frac{dp_i}{dt} &=& -\frac{1}{N}\sum^N_{j=1} \sin(q_i-q_j),
\end{eqnarray}
and temporal evolutions of the system are yielded from the equations
of motion. Each particle moves on the unit circle and interacts with
all the others. The $q_i$'s are phase of particles and the $p_i$'s are
canonical conjugated momenta. According to statistical mechanics, the
critical energy $E_c$ is $E_c/N=0.75$\cite{antoni-95}, where $N$
represents degrees of freedom. 

We define an order parameter parameter $M$ of this system as
\begin{equation}
  M = \frac{1}{L}\int^L_0 dt M(t)
  \label{order-parameter}
\end{equation}
and
\begin{equation}
  M(t) = ||\vec{M}(t)||,\quad
  \vec{M}(t) = \left( \frac{1}{N}\sum^N_{i=1}\cos q_i(t),
  \ \frac{1}{N}\sum^N_{i=1}\sin q_i(t) \right),
  \label{order-parameter-2}
\end{equation}
where $t$ and $||\vec{M}(t)||$ represent time and absolute value of
$\vec{M}(t)$, respectively. From the definition, $0 \leq M,\ M(t) \leq
1$. When all particles distribute uniformly on
the unit circle at each time, then $M=0$. On the other hand, when
particles form the cluster, which is the only one in this system
\cite{inagaki-93}, then $M\sim 1$.  For this system, Antoni and Ruffo
showed that energy dependence of order parameter obtained from
numerical experiments agrees with a result of statistical mechanics
\cite{antoni-95}.

We observe relaxation processes, that are temporal evolutions of
$M(t)$, toward equilibria using a local order parameter $M_{\tau}(n)$,
which is defined as
\begin{equation}
  M_{\tau}(n) = \frac{1}{\tau}\int^{n\tau}_{(n-1)\tau} dt M(t).
  \label{local-order-parameter}
\end{equation}

\begin{figure}[tbp]
  \begin{center}
    \leavevmode
    \epsfxsize=7cm
    \epsffile{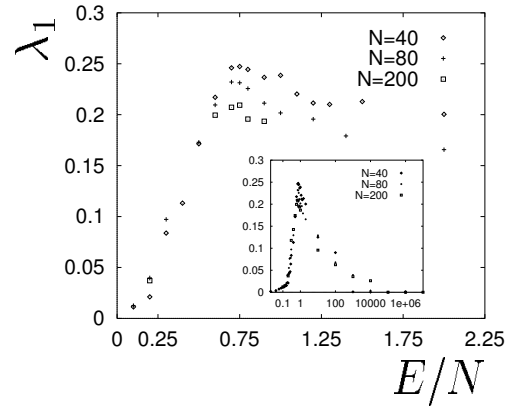}
  \end{center}
  \caption{The maximum Lyapunov exponent $\lambda_1$ vs. energy per
    unit particle $E/N$. $L=2^{23}\times 0.01$ ($\sim 10^5$, namely
    $10^7$ steps) (cf. Eq.(\protect\ref{order-parameter})).
    Inset: The horizontal axis is logarithm scale. $\tau=0.001,
    L=2^{20}\times 0.001$ when $E/N\geq 10^3$.
    ($\diamondsuit$): $N=40$. ($+$): $N=80$. ($\Box$): $N=200$.
    The quantity $\lambda_1$ takes the largest values at
    critical energy, namely $E_c/N=0.75$, and it goes to zeros in the
    limit of $E/N\to 0$ and $E/N\to\infty$.}
  \label{fig:Evslambda}
\end{figure}

The system Eq.(\ref{Hamiltonian}) is integrable in the limit $E/N\to
0$ and $E/N \to \infty$. For the former and the latter limit the
system is approximated as harmonic oscillators and free particles,
respectively. To explain the case of the latter limit, we show the
following expression of the potential term $U(q)$:
\begin{eqnarray}
  U(q)
  &=& \frac{N}{2}
  -\frac{1}{2N}\sum^N_{i,j=1}(\cos q_i\cos q_j+\sin q_i\sin q_j) \\
  &=& \frac{N}{2}(1-M(t)^2), \qquad 0 \leq M(t) \leq 1
\end{eqnarray}
(cf. Eq.(\ref{order-parameter-2})). Hence $U(q)$ has the minimum and
the maximum value 
\begin{equation}
  0 \leq U(q)/N \leq 1/2.
\end{equation}
We can therefore neglect the potential term when the value of $E/N$ is
large enough, then the system consists of free particles. Note that 
$\lambda_1\to 0$ when $E/N\to 0$ and $E/N\to\infty$ (see inset of
Fig.\ref{fig:Evslambda}).

To compare our numerical results with results of a phenomenological
theory of statistical mechanics, we introduce a phenomenological
theory: van Hove theory. We assume temporal evolution of order
parameter $M(t)$ is determined by gradient of Landau's free energy
$F(M)$, that is,
\begin{equation}
  \frac{dM(t)}{dt} \propto -\frac{\partial F(M)}{\partial M},
  \label{van-hove-eq}
\end{equation}
\begin{equation}
  F(M)=a'(T-T_c)M^2+b(T)M^4,\quad (a'>0).
  \label{landau-free-energy}
\end{equation}
Note that relaxations to equilibria occur at once if states are
non-equilibria ($\partial F(M)/\partial M\neq 0$). From the solution
of Eq.(\ref{van-hove-eq}), we obtain the following temporal evolution
of $M(t)$
\begin{equation}
  M(t)\sim\left\{
    \begin{array}{cc}
      \exp(-t/\tau(T)) & \quad(\mbox{when}\ T \glt T_c)\\
      t^{-1/2} & \quad(\mbox{when}\ T=T_c)
    \end{array}
  \right. ,
  \label{van-hove-result}
\end{equation}
where $\tau(T)$ is called relaxation time which depends on temperature
$T$ and diverges at critical temperature $T_c$, and
Eq.(\ref{van-hove-result}) is approximately correct for $t\gg\tau(T)$
when $T\glt T_c$. According to this theory, relaxation of power type
appears if and only if systems are just on critical points.

To integrate the equations of motion which are derived from the
Hamiltonian of Eq.(\ref{Hamiltonian}), we use forth order symplectic
integrators \cite{yoshida-93} with fixed time slice $\Delta t=0.01$
($\Delta t=0.001$ for a part of result in Fig.\ref{fig:Evslambda})
which keep symplectic properties of Hamiltonian systems exactly and
total energy accurately. Relative errors of total energy $\Delta E/E$
are less than $O(10^{-7})$ for considered energy region. The value of
$\tau$ in Eq.(\ref{local-order-parameter}) is fixed at $10$, namely
$1000$ steps except for the case given special comments. As initial
conditions we choose random variables which follow Gaussian
distribution. To see relaxation of order parameter, we take small
values for $q_i$'s and set $M\sim 1$. Scales of $p_i$'s are defined
from energy. In the model Eq.(\ref{Hamiltonian}) total momentum is
conserved, and we set total momentum is equal to zero.

This paper is organized as follows. In Section 2 results of numerical
experiments are reported. At first, slow relaxation of power type is
shown, then we investigate dependences on degrees of freedom and
initial condition for appearance of the slow relaxation. After that we
compare our numerical results with results of van Hove theory.  Next
we confirm that the system is highly chaotic at the critical energy
with the aid of temporal evolutions and probability distributions of
local order parameter. Section 3 is devoted to summary and
discussions.

\section{Results of simulations}
\label{sec:results}

In this section numerical results are reported for the Hamiltonian
system Eq.(\ref{Hamiltonian}). In Sec.\ref{sec:slow-relaxation} we
find that slow relaxation of power type appears at the critical energy
for a certain degrees of freedom and initial condition. We observe
dependences on degrees of freedom and initial condition for appearance
of the slow relaxation in Sec.\ref{sec:N-dependence} and
Sec.\ref{sec:initial-dependence}, respectively.
Section 2.4 is devoted to compare our dynamical results and results of
a phenomenological theory of statistical mechanics. After that we
investigate dynamical properties of the system, in particular, at the
critical energy. Energy dependence of maximum Lyapunov exponent is
shown in Sec.\ref{sec:lyapunov}. Then, in Sec.\ref{sec:distribution_lle}, we
confirm that the system is highly chaotic at the critical energy. That
is, the slow relaxation of power type may not be explained by theories
for perturbed systems. We study the structure of phase space in
Sec.\ref{sec:uniformity}, and suggest phase space is uniform at the
critical energy for the property of instability.

\subsection{Slow relaxation at the critical energy}
\label{sec:slow-relaxation}

\begin{figure}[bp]
  \begin{center}
    \leavevmode
    \epsfxsize=7cm
    \epsffile{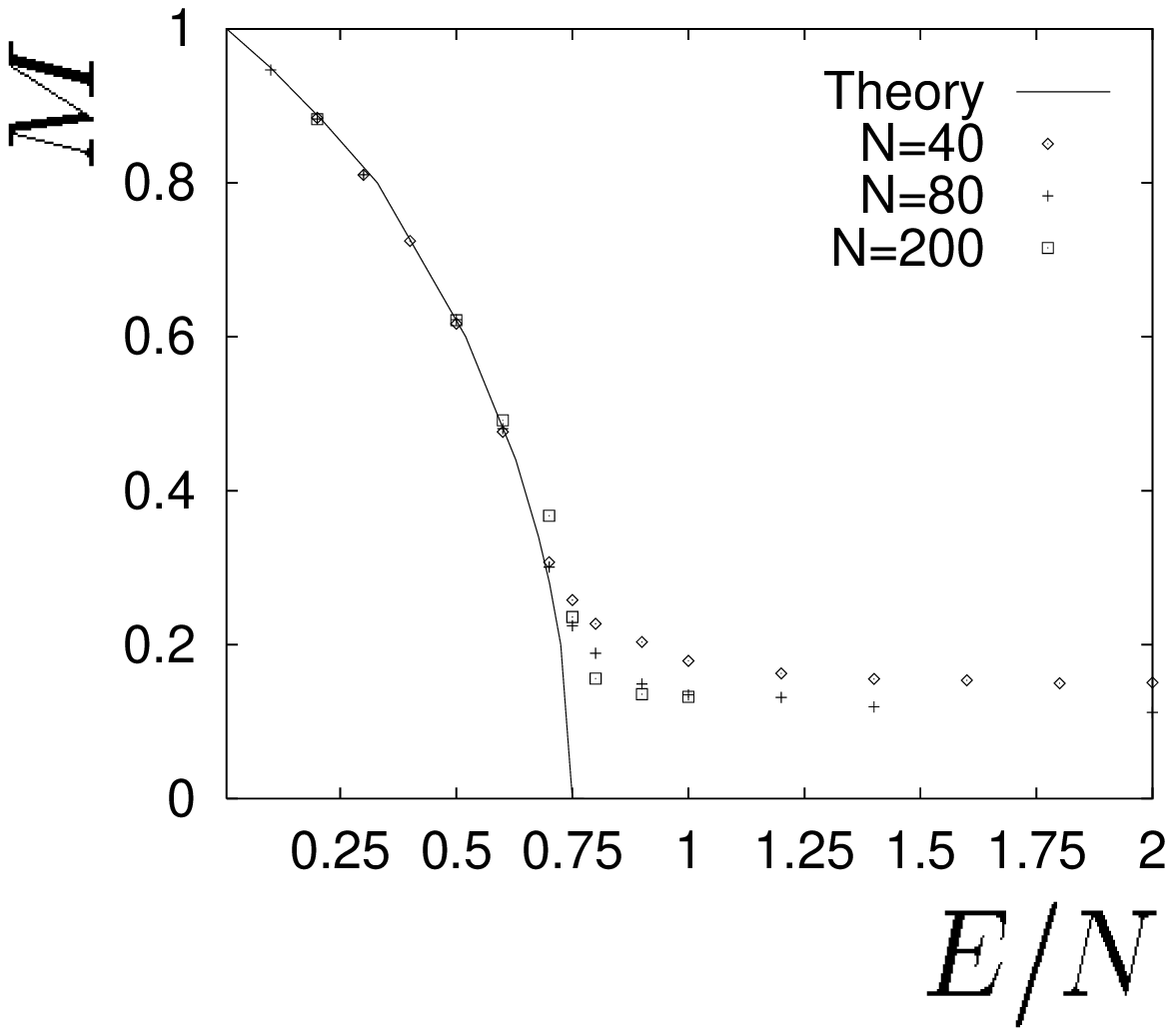}
  \end{center}
  \caption{Order parameter $M$ vs. energy per unit particle $E/N$.
    $L=2^{23}\times0.01$ ($\sim 10^5$, namely $10^7$ steps) (cf.
    Eq.(\protect\ref{order-parameter})). ($\diamondsuit$): $N=40$.
    ($+$): $N=80$. ($\Box$): $N=200$. Solid line represents a 
    result of theory of statistical mechanics. Results are good
    agreement with the theory.}
  \label{fig:EvsM}
\end{figure}

We show a energy dependence of order parameter $M$ in
Fig.\ref{fig:EvsM}. The solid line in the figure represents the curve
obtained from theory of statistical mechanics. Using the saddle point
method, it is described as simultaneous implicit functions
\cite{antoni-95}:
\begin{equation}
  M = \frac{I_1}{I_0}(\beta M),
\end{equation}
\begin{equation}
  E/N = \frac{1}{2\beta} + \frac{1}{2}(1-M^2),
\end{equation}
where $I_0$ and $I_1$ are Bessel functions of $0$-th and $1$-st order,
respectively. Then we find that results of numerical experiments fit
the theoretical curve, and the results are reproduction of results of
Antoni and Ruffo\cite{antoni-95}. Although we find differences
between the results and theory, we can understand causes of the
differences as follows. For high energy part ($E/N>1$), $M(t)$
fluctuates around zero, which is the minimum value of $M(t)$, since
the system is approximated as free particles, and $q_i(t)$'s takes
random variables. Then $M(t)$ is estimated around $O(1/\sqrt{N})$ from
central limit theorem. That is confirmed from $N$ dependence of the
values of $M$ in this part. For middle energy part ($E\sim E_c$), in
addition to the cause of differences in high energy part, there is
another cause. That is lack of time $L$ in which we take time average
of $M(t)$ because, according to van Hove theory, relaxation time
increases as energy goes to critical value.

\begin{figure}[hbtp]
  \begin{center}
    \leavevmode
    \epsfxsize=15cm
    \epsffile{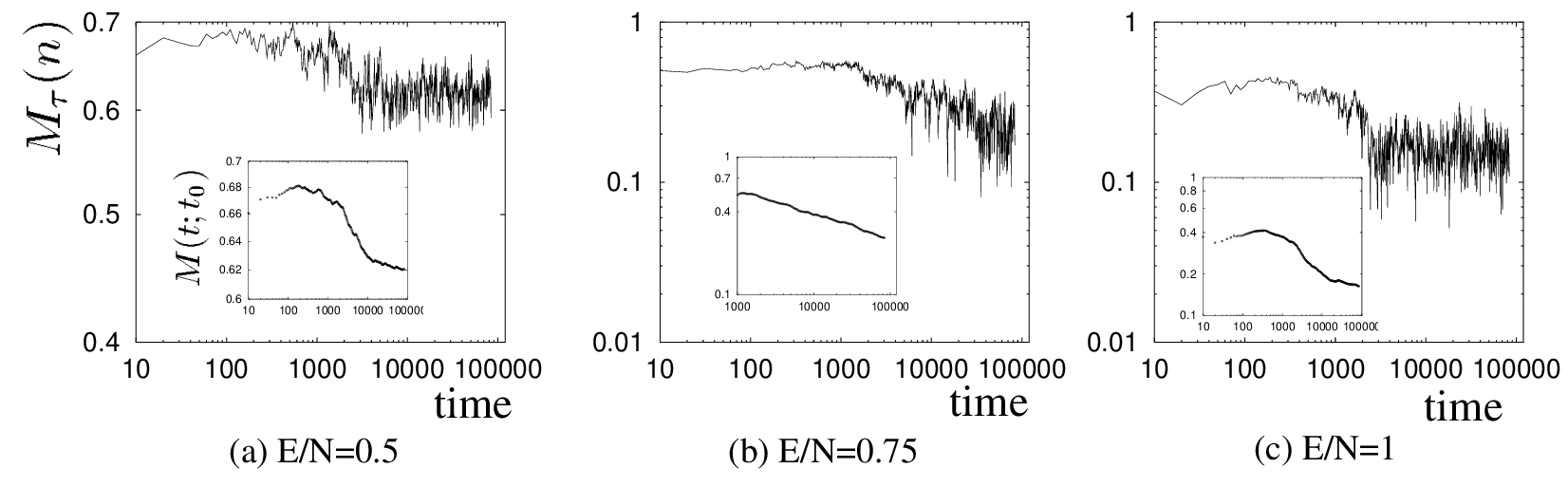}
  \end{center}
  \caption{Log-log plotted temporal evolutions of $M_{\tau}(n)$ (cf.
    Eq.(\protect\ref{local-order-parameter})). $N=80, \tau=10$. (a)
    $E/N=0.5$. (b) $E/N=E_c/N=0.75$. (c) $E/N=1$. Insets: The temporal
    evolution of $M(t;t_0)$ which is yielded by the time series of
    $M_{\tau}(n)$ (cf. Eq.(\protect\ref{convergence})), where
    $t_0=1000$ for (b) and $t_0=0$ for (a) and (c). When $E=E_c$ slow
    relaxation of power type appears from $t_0$.}
  \label{fig:tmp-evo}
\end{figure}

Here let us investigate relaxation of $M_{\tau}(n)$ defined as
Eq.(\ref{local-order-parameter}). We report results when $E/N=0.5,
0.75 (=E_c/N)$ and $1$ in Fig.\ref{fig:tmp-evo}. The insets of
Fig.\ref{fig:tmp-evo} represent the following quantity as a
function of time $t$ 
\begin{equation}
  M(t;t_0) = \frac{1}{t-t_0}\int^t_{t_0} dt M(t).
  \label{convergence}
\end{equation}
This quantity represents time average of $M(t)$ from $t_0$ to $t$, and
reduces fluctuation $\xi(t)$ when temporal evolution of $M(t)$ is
as follows
\begin{equation}
  M(t)= (t-t_0)^{-x}+\xi (t).  \end{equation} For
Fig.\ref{fig:tmp-evo}(b) we set $t_0=1000$ since slow relaxation of
$M_{\tau}(n)$ starts around the $t_0$. For Figs.\ref{fig:tmp-evo}(a)
and (c) we set $t_0=0$. From Fig.\ref{fig:tmp-evo}, we confirm the
relaxation is power type when $E=E_c$ for the initial condition. This
result agrees with a result of van Hove theory. The slow relaxation
finishes in finite time since $M(t)$ takes degree of $O(1/\sqrt{N})$
for $t\to\infty$.

\subsection{$N$ dependence of slow relaxation}
\label{sec:N-dependence}

We observed slow relaxation of power type when $N=80$. Now we show
that the slow relaxation appears even if we change degrees of freedom.
When $N=40$ the slow relaxation of power type is shown in
Fig.\ref{fig:N40-power}. Inset of the figure represents $M(t;t_0)$
(cf. Eq.(\ref{convergence}) against $t$ again, where we set
$t_0=5000$. However we have not observed the slow relaxation when
$N>80$, and we will discuss the reason and behaviors of the system in
thermodynamic limit in Sec.\ref{sec:summary}

\subsection{Initial condition dependence}
\label{sec:initial-dependence}

\begin{figure}[hbtp]
  \begin{center}
    \leavevmode
    \epsfxsize=15cm
    \epsffile{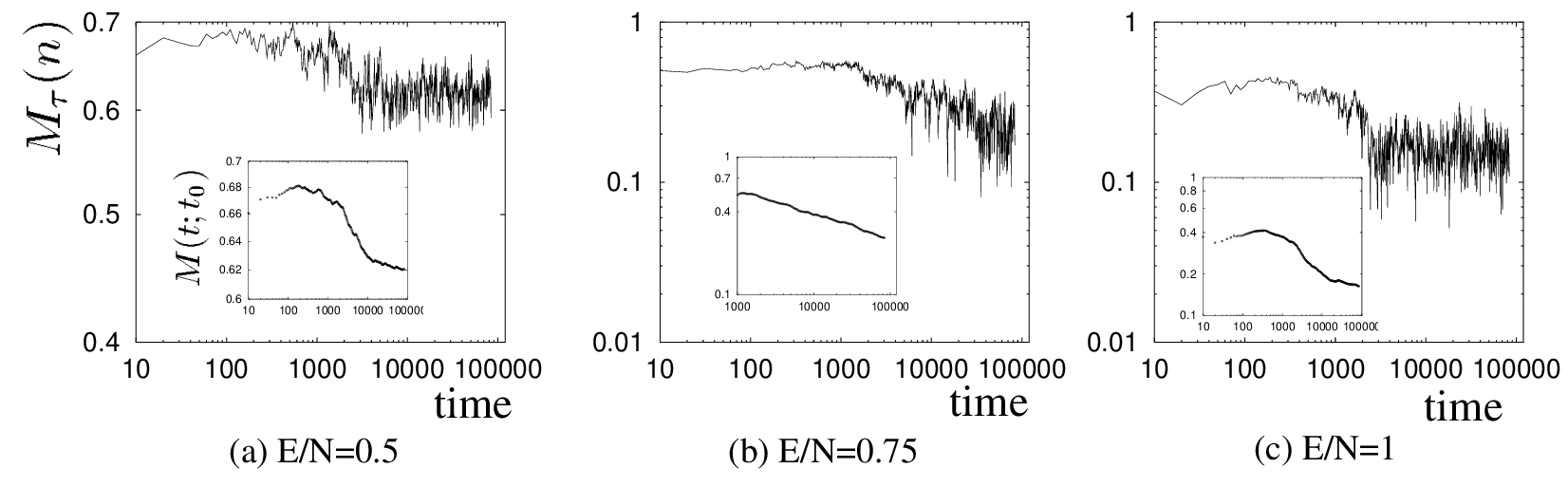}
  \end{center}
  \caption{Log-log plotted temporal evolutions of $M_{\tau}(n)$ (cf.
    Eq.(\protect\ref{local-order-parameter})) at $E_c$. $\tau=10$.
    Insets: Semi-log plotted graphs. (a) $N=80$. The initial condition
    is different from one of Fig.\protect\ref{fig:tmp-evo}(b). (b)
    $N=200$. (c) $N=1000$. We find relaxations are exponential type
    rather than power type.} 
  \label{fig:exp}
\end{figure}

In the previous section, we investigated dependence on degrees of
freedom $N$. Since we treat dynamical system we must study dependence
on initial conditions for relaxation of $M_{\tau}(n)$, too. A problem
is whether slow relaxation always appears or not when $E=E_c$. Figure
\ref{fig:exp} shows results when $N=80, 200$ and $1000$. Exponential
relaxations become clearer and clearer as $N$ increases hence slow
relaxation of power type does not always appear.  A rate of initial
conditions which yield slow relaxation may be low since only one
initial condition yields slow relaxation in ten initial conditions
when $E=E_c$ and $N=80$.  

\subsection{Comparing with a phenomenological theory}
\label{sec:comparing}

Up to now we observed appearance of slow relaxation of power type
at the critical energy for some initial conditions although the slow
relaxation does not always appears for any initial conditions even if
$E=E_c$. The appearance of the slow relaxation is a agreement with a
result of a phenomenological theory of statistical mechanics: van Hove
theory. However a disagreement also exists between our results and 
results of the phenomenological theory, and here we discuss the
disagreement. 

\begin{figure}[tbp]
  \begin{center}
    \leavevmode
    \epsfxsize=7cm
    \epsffile{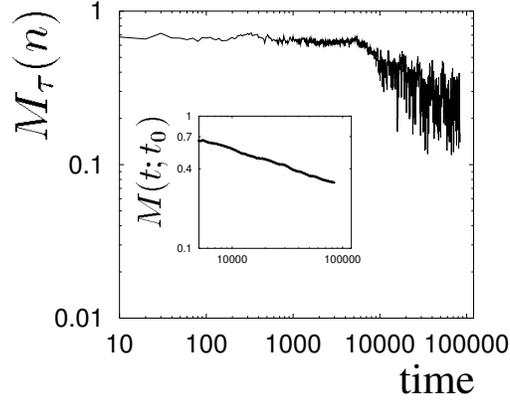}
  \end{center}
  \caption{The same with \protect\ref{fig:tmp-evo}(b), but $N=40,
    t_0=5000$.}
  \label{fig:N40-power}
\end{figure}

We found there are two time regions for slow relaxation process in
Figs.\ref{fig:tmp-evo}(b) and \ref{fig:N40-power}. One is the region
in which the system stays in non-equilibrium states (flat region), the
other is the region in which slow relaxation of power type occurs
(relaxation region). The existence of the flat region is different
from theory of van Hove, which says systems in non-equilibrium states
goes toward equilibrium states immediately (cf. 
Eq.(\ref{van-hove-eq})). Our dynamical results are therefore an
example of a disagreement with a result of van Hove theory.

\begin{figure}[bthp]
  \begin{center}
    \leavevmode
    \epsfxsize=7cm
    \epsffile{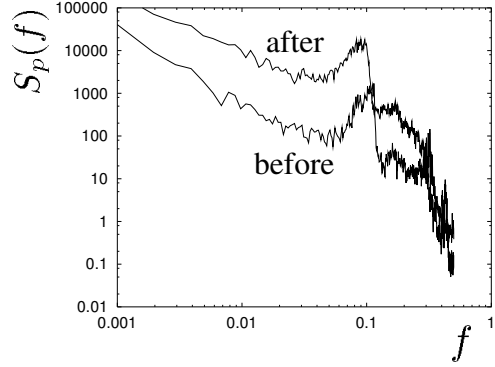}
  \end{center}
  \caption{Average of power spectra of momenta $p_j$'s
    $(j=1,2,\cdots,40)$ when $N=80$ and the initial condition yields
    slow relaxation of power type. Lower and upper graphs are
    calculated from time series before ($t\in [0,1024]$) and after
    ($t\in [2048,3072]$) the onset of slow relaxation, respectively.
    Distinguishing the two graphs, values of the vertical axis are
    multiplied $10$ times for the upper graph. The motion in phase
    space is not nearly periodic since the graphs do not consist of
    solitary peaks.}
  \label{fig:onset}
\end{figure}

The flat region means the existence of ``induction period'' and that
reminds induction phenomenon, which means that a nearly periodic
motion starts to behave as a non-periodic motion after a certain
amount of time, which is called induction period
\cite{hirooka-69}\cite{ooyama-69}\cite{saito-70}.
However the phenomenon appearing in Figs.\ref{fig:tmp-evo}(b) and
\ref{fig:N40-power} is different from induction phenomenon since the
motion in phase space is not nearly periodic both before and after the
onset of slow relaxation. To confirm that we show average of power
spectra of momenta $p_j$'s $(j=1,2,\cdots,N_0)$, $S_p(f)$, when $N=80$
in Fig.\ref{fig:onset}. The quantity $S_p(f)$ is defined as
\begin{equation}
  S_p(f)=\frac{1}{N_0}\sum^{N_0}_{j=1} S_j(f),
\end{equation}
\begin{equation}
  S_j(f) = \mbox{``power spectrum of time series of $p_j(t)$''}.
\end{equation}
Here we selected $N_0$ momenta from $N$ momenta to calculate $S_p(f)$,
and we set $N_0=40$. In low frequency region the power spectrum is
growing up hence the motion is not nearly periodic.

\subsection{Lyapunov exponent}
\label{sec:lyapunov}

We have investigated slow relaxation of power type using dynamical
method. Now we study dynamical properties of the slow relaxation and
structure of phase space in which the slow relaxation occurs.

As the first step, we investigate an energy dependence of the maximum
Lyapunov exponent $\lambda_1$ (hereafter we call this Lyapunov
exponent simply) which measures instability of orbits. Existence of
positive Lyapunov exponent means a sample orbit has instability. 
Results of numerical experiments are reported in
Fig.\ref{fig:Evslambda} and we find Lyapunov exponent takes the
largest value at $E_c/N(=0.75)$ for all sample orbits when $N=40, 80$
and $200$.

We can understand why value of Lyapunov exponent takes the maximum at
$E_c$ as follows. The system is integrable for low and high energy
limit since it is harmonic oscillators and free particles,
respectively. Thus $\lambda_1\to 0$ when $E/N\to 0$ and
$E/N\to\infty$. Integrabilities of those two limits break as energy
goes away from zero or infinity, and the two integrabilities balance
at the critical point. Consequently Lyapunov exponent is the maximum
at the critical point.

Butera et al. also investigated Lyapunov exponent for a
Hamiltonian system which is a two-dimensional system of coupled
rotators \cite{butera-87}. They found a ``knee-like'' shape in the
graph of Lyapunov exponent against temperature and the ``knee'' is at
the critical temperature of Kosterlitz-Thouless (KT) phase transition
\cite{kosterlitz-73} although the ``knee'' does not appear at the
largest value of Lyapunov exponent. Figure \ref{fig:Evslambda} shows a
``knee-like'' shape and the ``knee'' is at the critical energy where
$\lambda_1$ is the largest. This fact suggests that degree of
instability has some relations to phase transition. 

\subsection{Highly chaotic property at the critical energy}
\label{sec:distribution_lle}

We found that Lyapunov exponent takes the largest value at the
critical energy hence we expect the system is highly chaotic at the
critical energy. In this section we confirm that the system is highly
chaotic at the critical energy from temporal evolutions and
probability distributions of local Lyapunov exponents. The definition
of the local Lyapunov exponent is as follows
\begin{equation}
  \lambda_{\tau}(n) = \frac{1}{\tau}\int^{n\tau}_{(n-1)\tau} dt
  \lambda_1(t),
  \label{local-Lyapunov}
\end{equation}
\begin{equation}
  \lambda_1(t) = \frac{d}{dt} \log (\left|\left| X(t) \right|\right|)
  \label{lambda-t}
\end{equation}
where $X(t)$ is a $2N$-dimensional tangent vector at time $t$ which
obeys linearized equations of motion. The local Lyapunov exponent
$\lambda_{\tau}(n)$ indicates instability of orbits in the time
interval $[(n-1)\tau,\ n\tau]$. Orbits are instable if
$\lambda_{\tau}(n)>0$ in the time interval.

\begin{figure}[tbp]
  \begin{center}
    \leavevmode
    \begin{tabular}{cc}
      \epsfxsize=7cm
      \subfigure[$E/N=0.2$]
      {\epsffile{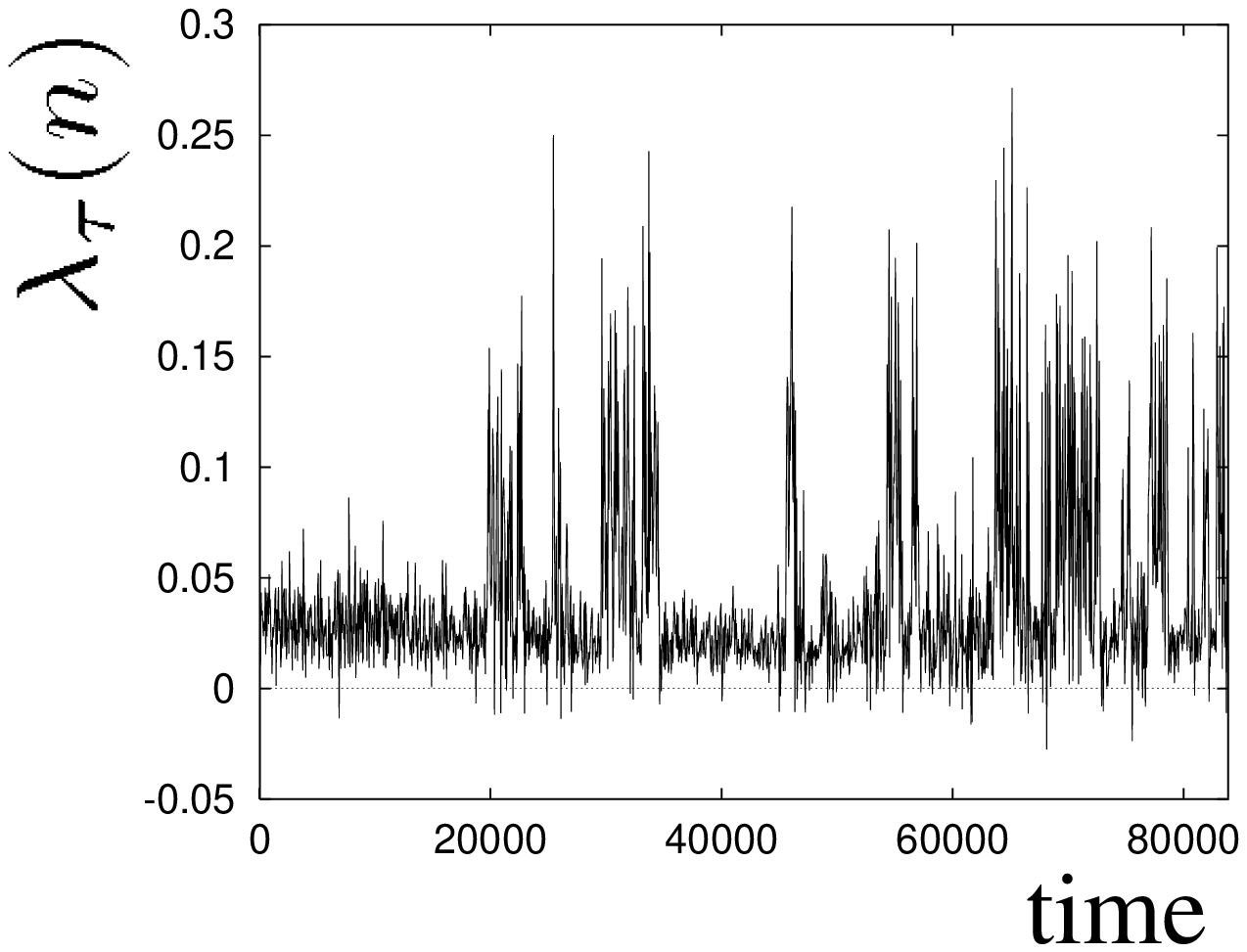}}
      \epsfxsize=7cm
      \subfigure[$E/N=0.75$]
      {\epsffile{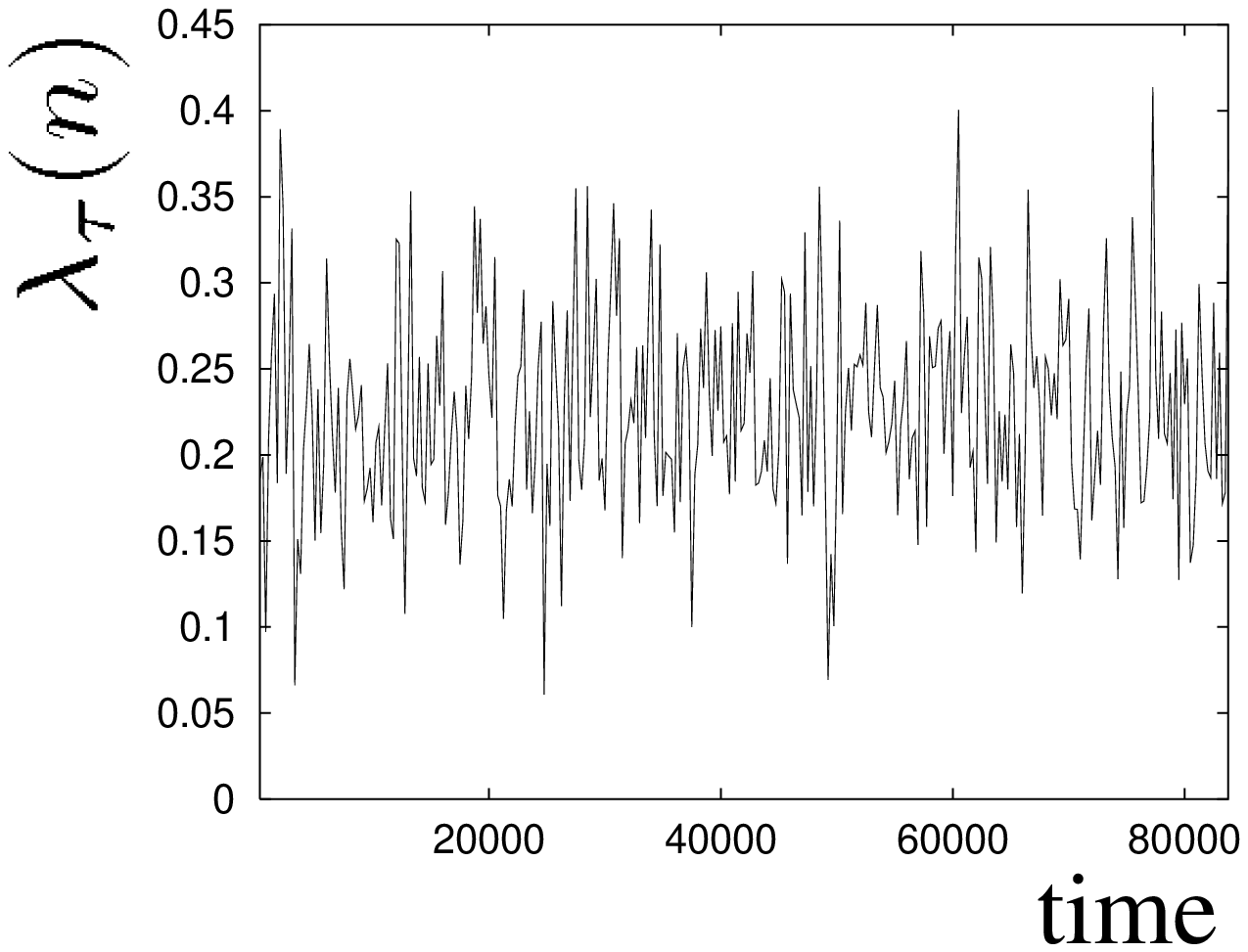}}
    \end{tabular}
  \end{center}
  \caption{Temporal evolutions of local Lyapunov exponents. $N=80$.
    (a): $E/N=0.2, \tau=50$. (b): $E/N=0.75=E_c/N, \tau=10$. The
    initial condition of (b) yields relaxation of power type. Typical
    intermittency is found in (a), and (b) is not intermittency.}
  \label{fig:lle-80deg}
\end{figure}

Figure \ref{fig:lle-80deg} shows temporal evolutions of local Lyapunov
exponents for $E/N=0.2$ and $0.75(=E_c/N)$. The initial condition of
Fig.\ref{fig:lle-80deg}(b) yields slow relaxation of power type (cf.
Fig.\ref{fig:tmp-evo}(b)). When $E/N=0.2$, typical intermittency is
found. On the other hand, when $E/N=0.75=E_c/N$, behavior of the
temporal evolution is different from the case of $E/N=0.2$ hence
motion is not intermittency for the initial condition. Moreover local
Lyapunov exponents do not take near zero, hence few tori, if any,
exist in phase space. Consequently the system is highly chaotic.

\begin{figure}[bthp]
  \begin{center}
    \leavevmode
    \epsfxsize=7cm
    \epsffile{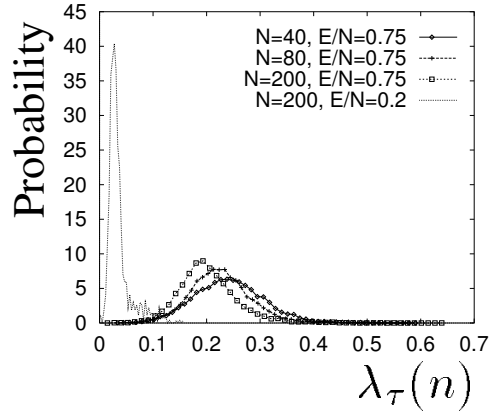}
  \end{center}
  \caption{Probability distributions of local Lyapunov exponents.
    ($\diamondsuit$): $N=40$. ($+$): $N=80$. ($\Box$): $N=200$. The
    value of energy is $E/N=0.75=E_c/N$. Dashed line is the case
    $N=200, E/N=0.2$, and intermittency occurs. We set $\tau=100$ when
    $E/N=0.2$. At the critical energy the system is different from
    intermittency.}
  \label{fig:distribution}
\end{figure}

To make sure of the difference between the two time series of local
Lyapunov exponent, we show probability distributions of local Lyapunov
exponents in Fig.\ref{fig:distribution}. If the system is weakly
chaotic and intermittency occurs, a probability distribution of local
Lyapunov exponents should have a peak near zero. However, according to
Fig.\ref{fig:distribution}, the distributions have not peaks near zero
at $E=E_c$. Thus we understand the system is different from weakly
chaotic system at $E=E_c$. Furthermore, at $E=E_c$, motion is not
reduced to Markovian process since relaxation of power type is
observed.

\subsection{Uniformity of phase space at the critical energy}
\label{sec:uniformity}

We observed slow relaxation of power type in highly chaotic system. If
the system is weakly chaotic, we can understand the cause of the slow
relaxation from the structure of phase space using theories for
perturbed systems. However the slow relaxation occurs in highly
chaotic system, thus we can not use the theories and we must
newly investigate the structure of phase space to understand the cause
of the slow relaxation. Here we suggest that phase space is uniform at
the critical energy. 

For the purpose, we calculate Lyapunov exponent for ten initial
conditions when $E=E_c$ and $N=80$. Results are as follows;
\begin{eqnarray}
  \lambda_1 &=&  0.229331,\ 0.229133,\ 0.231313,\ 0.229381,\
  0.229121, \nonumber \\ 
  && 0.231387,\ 0.230692,\ 0.231396,\ 0.228699,\ 0.230357.
  \label{lambdas}
\end{eqnarray}
That is,
\begin{equation}
  \lambda_1 = 0.2300 \pm 0.0014 \quad (E/N=0.75, N=80).
  \label{lambda-initial}
\end{equation}
%
\begin{wraptable}{r}{7cm}
  \begin{center}
    \leavevmode
    \begin{tabular}{cc}
      $N$ & standard deviation \\
      \hline
      40 & 0.065916 \\
      80 & 0.056067 \\
      200 & 0.055623
    \end{tabular}
  \end{center}
  \caption{Standard deviations of local Lyapunov exponents when
    $E/N=E_c/N=0.75$. Those probability distributions are described in 
    Fig.\protect\ref{fig:distribution}.}
  \label{tab:stand-dev}
\end{wraptable}
On the other hand, Table \ref{tab:stand-dev} shows standard deviations
of local Lyapunov exponents for the critical energy $E_c/N=0.75$
described in Fig.\ref{fig:distribution}. Then we find that the
distribution of Lyapunov exponents is narrow enough comparing with the
standard deviations of local Lyapunov exponents. Hence initial
conditions have no influences on the values of $\lambda_1$, in other
words, phase space is uniform for the intensity of instability at the
critical energy.

The values of $\lambda_1$ are not different between relaxations of
power type and exponential type because the first and the second value
of Eq.(\ref{lambdas}) correspond to the initial conditions of
Figs.\ref{fig:tmp-evo}(b) and \ref{fig:exp}(a), respectively. Hence we
can not distinguish between the two types only from the values of
$\lambda_1$, which are time average of instability.

\section{Summary and discussions}
\label{sec:summary}

Relaxation processes are numerically investigated through temporal
evolutions of local order parameter in a Hamiltonian system which has
second order phase transition. Slow relaxations of power type of
order parameter are observed at the critical point for some initial
conditions. It is the first time that power type decay is observed for
order parameter, which is a important quantity to observe the system,
from temporal evolutions of equations of motion derived from a
Hamiltonian.

To understand the cause of the slow relaxation dynamically, we
investigated dynamical properties of the system. Then we found that 
the slow relaxations occur in highly chaotic systems rather than
weakly chaotic systems. That is, mechanism of the slow relaxations can
not be explained by theories for perturbed systems, hence we must
consider new theories for highly chaotic systems.

The slow relaxations of power type are observed when $N=40$ and $80$.
However we have not observed the slow relaxation when $N>80$, and some
initial conditions yield exponential relaxations even if $E=E_c$.

Why the slow relaxation has not been observed when $N$ is large?
Large number of $N$ makes a situation such that to detect slow
relaxation is difficult because the larger $N$ is, the clearer the
critical point may be, and energy has small errors in calculations
although we set $E=E_c$ initially. Moreover we do not know the exact
value of the critical energy in the meaning of dynamics. In other
words, we do not know whether dynamics also has the same critical
energy $E_c$ which is obtained from statistical mechanics.
However to obtain exact value of critical energy is difficult since
appearance of the slow relaxation depends on initial conditions. We
can not say whether the slow relaxation appears in the thermodynamic
limit. The author expects the appearance because we have supporting
evidences for appearance but do not have for absence.

Appearance of the slow relaxation agrees with a result of
phenomenological theory of statistical mechanics; van Hove theory.
However, contrary to the phenomenological theory, relaxation processes
do not start at once even the system are in non-equilibrium states. 
That is, we observed an agreement and a disagreement between our 
results from numerical experiments and results from a phenomenological
theory. To research the origin of the agreement is interesting to
understand contracted dynamics which yields motion of order parameter.

From temporal evolutions and probability distributions of local
Lyapunov exponents, we understood the system is highly chaotic rather
than weakly chaotic at the critical energy. Hence we may not apply
theories for perturbed systems to understand the slow relaxation of
power type at the critical energy. We must make new theories for
highly chaotic systems. For the purpose we must understand structure
of phase space when the system is highly chaotic and has many degrees
of freedom. To understand the structure of phase space, a result is
obtained such that the structure of phase space may be uniform for the
intensity of instability at the critical energy.\\
\ \\
\ \\ 
{\Large\bf Acknowledgement}\\ 
\ \\ 
I express my thanks to Tetsuro Konishi for useful discussions and a
careful reading of the manuscript. I acknowledge helpful discussions
with Hiroyasu Yamada and Akira Yoshimori. I wish to thank
Kazuhiro Nozaki for special encouragements. I thank referees for
fruitful comments.

\newpage

\newpage

\end{document}